\documentclass{ws-ijmpe}
\usepackage[super,compress]{cite}
\usepackage{url}
\usepackage{amsmath,amssymb,amsfonts}
\usepackage{bm}
\usepackage{verbatim}
\usepackage{hyperref}
\usepackage{slashed,braket}

\newcommand{\tr}{\operatorname{tr}}
\newcommand{\cA}{\mathcal{A}}
\newcommand{\cD}{\mathcal{D}}
\newcommand{\cF}{\mathcal{F}}
\newcommand{\D}{\Delta}

\begin{document}


\title{Non-perturbative suppression of Chiral Effects in hot QGP for \(\Lambda/\bar\Lambda\)-hyperons spin polarization in heavy-ion collisions}

\author{Ruslan A.~Abramchuk}

\address{Physics Department, Ariel University\\
Ariel 40700, Israel\\
ruslanab@ariel.ac.il, abramchukrusl@gmail.com}

\maketitle

\begin{history}
\accepted{15 Oct 2024}
\end{history}

\begin{abstract}
    We review the calculation of suppression of Chiral Separational  
    and Chiral Vortical Effects  for strange quarks 
        (which allegedly yield spin polarization of \(\Lambda/\bar\Lambda\)-hyperons in peripheral Heavy-Ion Collisions)
    by the non-perturbative interactions in hot deconfined QCD
    with the Field Correlator Method. 
    The parameter range in the temperature-baryon density plane 
    is expected to cover LHC-ALICE and RHIC-STAR data.
\end{abstract}


\ccode{PACS numbers: 12.38.-t, 12.38.Mh, 12.40.-y, 25.75.-q}


\section{Introduction}\label{SectIntro}

\begin{figure}
    \center{
    \includegraphics[width=0.48\textwidth]{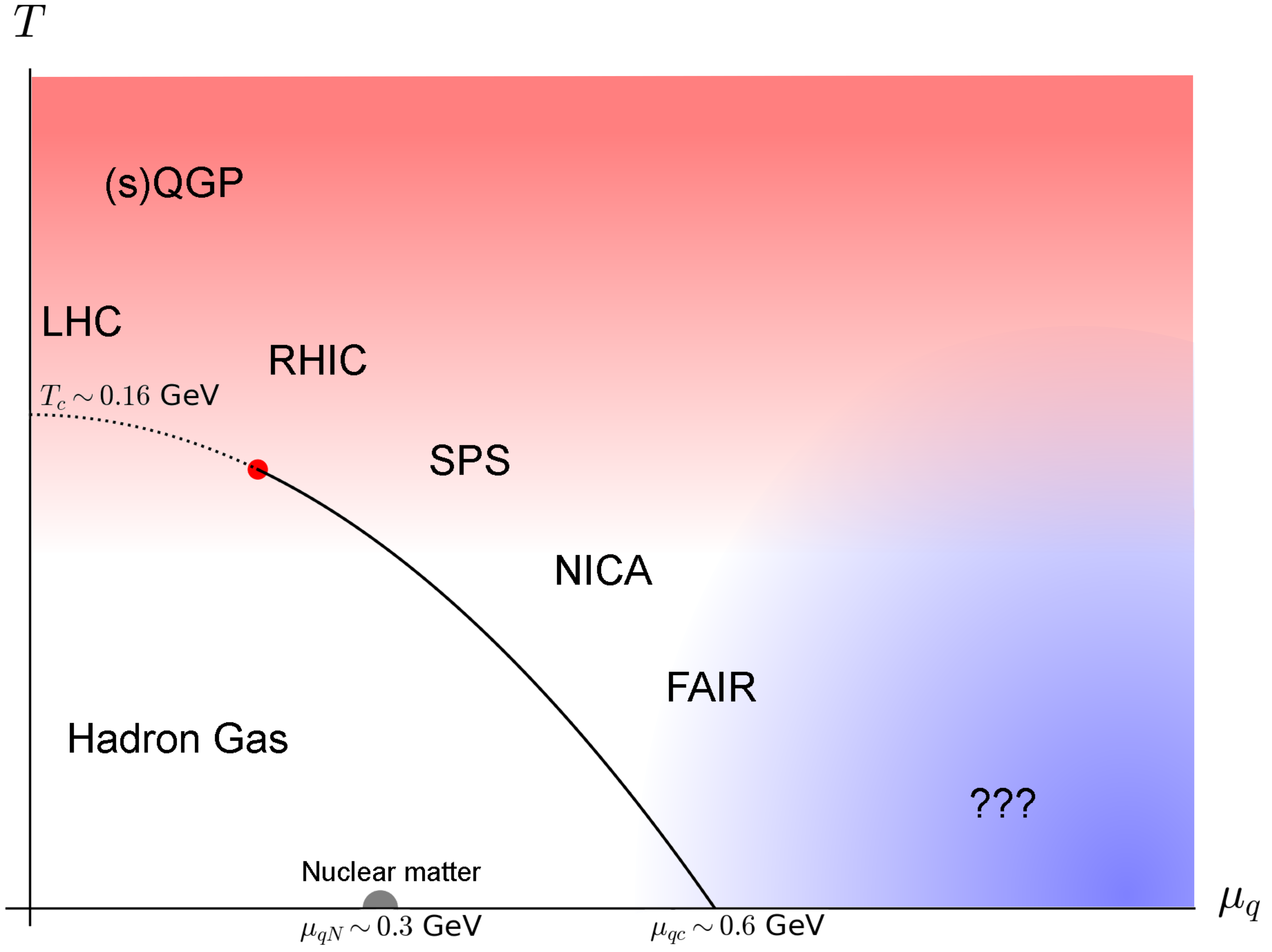}
    }
    \caption{
        QCD phase diagram\cite{QCDphases,1,2,3,5,7,8,10,Krivoruchenko2010jz} sketch. 
        Quark chemical potential \(\mu_q\) is related to the baryon chemical potential \(\mu_B\) as \(\mu_B=3\mu_q,~\mu_q=\mu_u=\mu_d,~\mu_s,\mu_I=0\).
    \label{FigPhaseDiagram}
    }
\end{figure}

Extreme conditions, 
    that emerge in modern collider experiments on ultra-relativistic collisions of heavy ions,
are unattainable in any other process in the Universe 
    (except may be for the conditions in the very early universe). 
The emerging matter --- Quark Gluon Plasma (QGP), thermalizes 
and reaches extremely high temperature, 
    but behaves like an ideal fluid, rather than like an ideal gas\cite{STAR2005gfr}.
With the unprecedentedly small viscosity to entropy density ratio\cite{Sharma2009zt} \(\eta/s\sim1/4\pi\), 
    QGP is the most perfect fluid known.
QGP is also probably the most opaque matter known,
    since the microscopical quantity of the matter absorbs high energy jets,
    that is suggested by the jet-quenching\cite{ALICE2015mdb}.
In non-central collisions, the huge (per nucleon) mutual angular momentum of the ions 
transforms into extremely high vorticity\cite{STAR2017ckg} of the QGP, 
    and the electric currents produce extremely high magnetic fields\cite{Bzdak2011yy}.
In QGP, the strong interactions dominate, 
    so hot dense QCD (Fig.\ref{FigPhaseDiagram}) is a natural model.

A soft probe for the vorticity and magnetic fields 
    in peripheral Heavy-Ion Collisions (HIC)  
    is the \(\Lambda/\bar\Lambda\)-hyperons spin polarization\cite{STAR2017ckg}.
There are two (competing or complimentary, which is a separate question) approaches to the polarization phenomena.
A thermal vorticity approach, 
    based on the assumption of local thermodynamical equilibrium of weakly-interacting hyperons right after the QGP freeze-out,
    suggested in \refcite{Becattini2013vja,Becattini2016gvu} allows to model\cite{Karpenko2016jyx,Karpenko2017lyj,Bravina2021arj} the experimental data\cite{STAR2017ckg}.

Another, quark approach,
    based on an assumption that the spin polarization of quarks in QGP freezes out into the spin polarization of baryons,
    suggested in \refcite{Rogachevsky2010ys,Baznat2013zx,Baznat2015eca,Sorin2016smp} for the s-quark\cite{Sorin2016smp} observed as the 
\(\Lambda/\bar\Lambda\)-hyperons polarization also allows to model\cite{Baznat2017jfj} the experimental data\cite{STAR2017ckg},
    including a native explanation\cite{Baznat2017jfj,Sorin2016smp} of the high \(\bar\Lambda\) (in comparison to \(\Lambda\)) polarization at relatively low energies \(\sqrt s_{NN}\lesssim 10\) GeV
    (while the feature might be an issue in the thermal vorticity-based approach\cite{Karpenko2017lyj,Bravina2021arj}).
    
In this paper we adopt the quark-based approach, 
    since the approach involves Chiral Effects as a part of the mechanism of the observed polarization.
Qualitatively, since the initial state of the system of colliding ions contains only nucleons, 
    the net strangeness of the system is zero at any moment
        (up to absolutely negligible effects of the weak interaction).
However, the strange quarks pairs production cross-section \(q\bar q\to s\bar s,~gg\to s\bar s\) 
is considerable during the collision and the QGP evolution,
    so there are fluctuations of the strangeness density in the fireball.
Vorticity and magnetic field interact with the strange (anti)quarks, and
    lead to P-odd effects\cite{Baznat2013zx,Baznat2015eca,Sorin2016smp,Baznat2017jfj,Zinchenko2022tyg}. 

In particular, the hyperons covariant polarization pseudovector \(\Pi_\mu\) 
(in the laboratory frame or in the hyperon rest frame) is related (see \refcite{Sorin2016smp} for a derivation) to the axial charge of strange quarks in QGP \(Q_5^s\) as
\begin{gather} 
    \Pi^{\Lambda,\text{lab}} = \frac{\Pi^\Lambda_0}{m_\Lambda}(p_y,0,p_0,0),\quad
    \braket{\Pi_0^\Lambda} 
        = \braket{\frac{m_\Lambda}{N_\Lambda p_y}} Q_5^s,
\end{gather}
where \(N_\Lambda\) is the number of hyperons with transverse to the reaction plane momentum component \(p_y\), 
    which is parallel to the total angular momentum of the system, 
    and \(m_\Lambda=1116\) MeV is the hyperons mass.

The axial charge is allegedly\cite{Sorin2016smp,Baznat2015eca} acquired by the strange quarks via the Chiral Vortical (CVE) and Separational (CSE) Effects.
Effect of magnetic field, and hence CSE, is at least an order of magnitude weaker\cite{STAR2017ckg} 
    than effect of vorticity
    (yet CSE as a correction to CVE might be observed in a comparative study of isobars collisions,
    e.g. \( ^{96}_{44}Ru\) and \( ^{96}_{40}Zr\) at RHIC).
The dominant CVE-based contribution reads\cite{Sorin2016smp}
\begin{gather}
    Q_5^s=\int d^3x~N_cc_V(\mu_s(x), T(x))\gamma^2\epsilon^{ijk}v_i\partial_jv_k,
\end{gather}
    \(v\) is the hydrodynamic velocity of the QGP, \(\gamma = (1-v^2)^{-1/2}\);
\(\mu_s\) is the strange quarks chemical potential and \(T\) is the temperature of QGP at a given point.
\(c_V(\mu_s(x), T(x))\) is the coefficient in CVE, 
    which in a conventional form is represented\cite{VilenkinCVE} as chiral current proportionality to vorticity \(\vec j_5 = c_V~\vec\omega\).

In the simulations \refcite{Baznat2017jfj} the free value \(k=1\) of the CVE coefficient 
\begin{equation}
    c_V(k)=k\frac{T^2}{6}+\frac{\mu^2}{2\pi^2}, \label{EqCvk}
\end{equation}
    would led to overestimation\cite{Sorin2016smp} of the result by an order of magnitude in comparison to the signal found in the RHIC-STAR data\cite{STAR2017ckg}.
The suppression of CVE  was attributed\cite{Sorin2016smp} to the correlation effects in QGP, 
    and accounted for by setting\cite{Baznat2017jfj} the suppression factor \(0\le k<1\).

In this paper, we present an analytic assessment\cite{Abramchuk2023} of this suppression at the stage of QGP evolution. 
Additional corrections may emerge during the freeze-out, which are yet to be investigated.
The leading non-perturbative interactions in QGP are accounted for with the Field Correlator Method\cite{Simonov2018cbk,Simonov2002vva,Simonov2009zx,Simonov2007jb,Krivoruchenko2010jz,Abramchuk2019,OrlovskySimo,Agasian2017,Andreichikov2017ncy,Agasian2006ra,Simonov2016xaf} (FCM).
The parameter range is expected to cover LHC-ALICE and RHIC-STAR data, see Fig. \ref{FigPhaseDiagram}
--- temperatures above the crossover transition \(T_c\sim 160\) MeV 
    and relatively small quark densities 
        (at least less than the nuclear density).
SPS, NICA and FAIR might require a separate assessment.

Perturbative corrections to CVE have been analyzed in more general theoretical context in \refcite{Golkar2012kb}.
Enhancement for the thermal part of CVE was proposed from a perturbative two-loop calculation in \refcite{Hou2012}.
Suppression was suggested in \refcite{Buzzegoli2021jeh},
    and estimated as \(\sim 40\%\) for the temperature range \(175-300\) MeV,
        which is in a qualitative agreement with our result (Fig.\ref{FigICVE}).
Calculation of the leading radiative correction to CSE in QED was attempted in \refcite{Gorbar2013upa}
CSE was studied with lattice Monte-Carlo simulations in QCD in \refcite{Brandt2022,Brandt2023,Buividovich2013hza}, 
    which is also in an agreement with our result (Fig.\ref{FigCSE});
    a similar result was obtained\cite{Khaidukov2023pfu} in the strong magnetic field limit.
CVE study with lattice simulations is possible, 
  but technically challenging. 

This paper recaps our recent results on CSE\cite{Zubkov2023} and CVE\cite{Abramchuk2023}.
In the next Section \ref{SectFCM} we review FCM application to high temperature QCD.
    In Section \ref{SectCSE} modification of CSE in hot QCD is considered, 
    and in Section \ref{SectCVE} --- of CVE (rigidly rotating hot QCD as the semi-realistic model for vortical QGP).
    Then we visualize and discuss the results in the Section \ref{SectRes}, 
    and make a brief conclusion in the Section \ref{SectCon}.

\section{FCM quark propagator in hot QGP}\label{SectFCM}

The key point of the Field Correlator Method 
    (see \refcite{Simonov2018cbk} and references therein) 
is decomposition of the gluonic field \(A_\mu\) in the background \(B_\mu\) and perturbative \(a_\mu\) parts
\begin{gather*} 
  A_\mu=B_\mu+a_\mu,\quad
 \cD A=\cD B\cD a.\,
\end{gather*}
The perturbative part is defined according to the well-established Background Perturbation Theory.
  The background part 
  (in the Matsubara representation, \((\vec x,\tau)\in\mathbb{R}_3\times S_1\), with the \(O(4)\) Euclidean symmetry explicitly broken) 
    is defined via five gauge invariant correlators \(D^{E,H},\,D_1^{E,H,EH}\). 
E.g., the spatial component of the background field strength tensors 
    (\(F_{\mu\nu} = \partial_\mu B_\nu - \partial_\nu B_\mu + [B_\mu,B_\nu]\)) 
    correlate as 
\begin{align} 
    N_c^{-1}\tr_c\braket{F_{ij}(x)\Phi_{xy}F_{kl}(y)\Phi_{yx}}_B 
        &= (\delta_{ik}\delta_{jl}-\delta_{il}\delta_{jk})D^H + \nonumber\\
    + \frac12\partial_i&((((y-x)_k\delta_{jl}-(y-x)_l\delta_{jk})+\text{perm})D_1^H).
\end{align}
The gauge invariant correlators are to be defined from self-consistency conditions.
The background field \(B_\mu\) is conjectured to be stochastic, 
which agrees with lattice data 
(and was proven to be self-consistent at zero temperature and baryon density).
Lattice data suggest\cite{DElia1997,DElia2002} that the correlators are short-range (in the Euclidean space),
with the correlation length weakly dependent on temperature
\(D(x,y)\sim e^{-|y-x|/\lambda},\,\lambda \sim 1\text{ GeV}^{-1}\sim 0.2\) fm.
The deconfinement corresponds to vanishing of the scalar Color-Electric correlator (\(D^E\to 0\)),
which occurs at \(T_c\sim 160\) MeV at \(\mu_q=0\).

The method also relies on an expansion,
    but instead of the perturbative expansion in powers of \(\alpha_s\),
        the non-perturbative cluster expansion in powers of a 
            small parameter \(\sigma\lambda^2\sim1/5\),
            which is defined eventually by the gauge group algebraic structure
                (\(\sigma\) is the confining string tension at zero temperature).
E.g., for a stochastic zero-mean variable \(\xi(t)\) the cluster decomposition 
    in Gaussian approximation reads
\(
    \braket{\exp(-i\int\xi(t)dt)} =
        \exp\Big(\sum_n \frac{(-i)^n}{n!}\int dt_1\ldots 
        dt_n\braket{\braket{\xi(t_1)\ldots\xi(t_n)}}\Big) 
    \approx \exp\Big(-\frac12\iint dt_1dt_2\braket{\braket{\xi(t_1)\xi(t_2)}}\Big).
\)

The quark propagator in gluonic background reads in the world-line representation 
\begin{align}
    S(x,y) =& (-\slashed D(A, \mu) + m)_{x} 
        \int_0^{+\infty}ds~\zeta(s)~(\overline{\cD^4z})_{xy}^s e^{-m^2s-K}\times\nonumber\\
    &\times P\exp\Big(ig\oint A\cdot dz 
        + \mu\int dz_4 
        +\int_0^sd\tau\sigma^{\rho\sigma}gF_{\rho\sigma}[A](z,z_0)\Big) \label{EqSq}.
\end{align}
The path integral \((\overline{\cD^4z})_{xy}^s\) describes the fermion motion from point \(x\) to \(y\) in world-line (proper) time \(s\).
The bar indicates the anti-periodic boundary conditions for the fermion 
    as the trajectory wraps \(n\) times around the temporal direction
        in the Euclidean space \(\mathbb{R}^3\times S^1\) with the compactified temporal direction 
        of length \(\beta=T^{-1}\).

In a gauge invariant operator,
    which constitutes any observable in QCD
        including axial current density,
    the Schwinger phase factors necessarily add up to closed Wilson loops \(W[{\cal C}]\).
Then the average over the background gluonic field \(B\) is to be calculated.

Following \refcite{Abramchuk2023} and references therein,
we neglect correlation between the (\(n\)-times winded up) Polyakov line \(L^{(n)}\),
    which is a temporal projection,
and the spatial projection \(W_3[{\cal C}]\) of the Wilson loop 
\begin{gather}
    \braket{W[{\cal C}]}_B = \braket{\exp\left(ig \oint_{\cal C} {\cal B}_\mu dz^\mu  \right)}_B 
        \approx \braket{ W_3[{\cal C}]}_B L^{(n)}.
\end{gather}
The algebraic part of the factorization problem may be resolved with the Diakonov-Petrov approach\cite{Zubkov2019ity}.

The Polyakov line  
is approximately parametrized with the potential \(V_1\) as
\(
  L^{(n)} \approx L^{|n|}\approx\exp\left(-\frac{nV_1(T)}{2T}\right)
\).
\(V_1(T)=V_1(r\to +\infty,T)\) is the singlet free energy of quark-antiquark white state,
is apparently produced by the vector Color-Electric correlator \(D_1^E\).
A fit\cite{Simonov2007jb} of Lattice QCD data for \(L\) is used as input for \(V_1(T)\).

The spatial projection \(W[{\cal C}_l]\) of the Wilson loop \(W[{\cal C}]\) obeys the area law due to the Color-Magnetic Confinement (CMC)\cite{Agasian2006ra},
    which generates the Debye (thermal screening) mass of the quark, 
\begin{gather}
    \braket{W_3[{\cal C}]}_B \approx \exp(-\sigma_H S_3[{\cal C}_l]),\quad
    \sigma_{H,f}(T)\approx c_\sigma^2g^4(T,\mu)T^2,\quad \nonumber\\
    M = \sqrt{m^2 + (c_D^2/4-2/\pi)\sigma_{H,f}(T)},\quad
\end{gather}
The coefficients \(c_\sigma\approx 0.56,~c_D\approx 2.0.\) were calculated\cite{Agasian2023bhe} within FCM,
    in a quantitative agreement with lattice QCD data. 

Noteworthy that the spin-CMC interaction 
    (the spin-spin contribution to the non-perturbative quark self-energy) 
considerably shifts down the effective quark mass
\begin{align}
    \D m^2 = -\Lambda &= 
        -\int d^4(y-x) \braket{\sigma^{ij}F_{ij}(x)\Phi_{xy}\sigma^{kl}F_{kl}(y)\Phi_{yx}}_B G(x,y)
        \nonumber\\
    &\approx -\frac2\pi\sigma_{H,f}(T).
\end{align}

In this paper, only the `kinematic' effects of the quark chemical potential are embodied. 
At large densities (\(\mu\sim\mu_{qc}\sim 0.6\) GeV the Field Correlators are less studied \cite{Krivoruchenko2010jz,Simonov2007jb}, 
    so there is space for phenomenology ---
we expect the CMC and the Color-Electric interaction to be additionally suppressed\cite{Krivoruchenko2010jz}.
The analysis is additionally complicated in isospin-asymmetric medium\cite{Avdoshkin2017cqp},
but we concentrate on the strange quark flavor in isospin-symmetric medium.

\section{Chiral Separation Effect in hot QGP}\label{SectCSE}

    For calculation\cite{Zubkov2023} of the axial current,
      we use the quark propagator \eqref{EqSq} in external magnetic field
      (where all the perturbative corrections are neglected)
    \begin{align}
      &\braket{j^5_\mu(x)} \approx \langle \tr_{c,D} \gamma_5\gamma_\mu (-\slashed D(B, \cA) + m)_{x} 
      \int_0^{+\infty}ds~\xi(s)~(\overline{\cD^4z})_{xx}^s e^{-m^2s-K} \times\nonumber \\
      &\times P\exp\left(ig\oint B\cdot dz + iq\oint\cA\cdot dz
      + \int_0^sd\tau\sigma^{\rho\sigma}(gF_{\rho\sigma}(z,z_0) + q\cF_{\rho\sigma})\right)\rangle_B,
    \end{align}
    where the spin coupling to the magnetic field is provided by the last term in the Schwinger phase factor.
    The quark chemical potential \(\mu\) is absorbed in the definition of the external electromagnetic field \(\cA\), 
    so \(q\oint\cA\cdot dz^{(n)} = q\Phi_H -i\mu n \beta\).

    Then the leading non-perturbative contribution to CSE reads 
    (with the Fermi distribution \(f(\varepsilon) = (e^{\beta\varepsilon}+1)^{-1}\))
    \begin{align}    
        &\braket{j^5_z}  \approx \frac{N_cq^2H}{2\pi^2} I_\text{CSE},
        \quad \sigma_{\text{CSE}}(T) =\frac{1}{2\pi^2}\left.\frac{d I_\text{CSE}}{d\mu}\right|_{\mu=0},\\
        I_\text{CSE} = -\int_0^{+\infty}&\frac{p^2dp}{\sqrt{{M^2}+p^2}}
            \left(f'(\sqrt{p^2+{M^2}} +{ V_1/2} -\mu) - (\mu\to-\mu)\right),
            \label{EqICSE}\\
        &\to \mu, \text{ at }V_1,M\to 0.
    \end{align}
    with the correct free massless limit.

    At large \(\mu\) the obtained result \eqref{EqICSE} approaches the free limit\cite{Zubkov2023},
        though the result is qualitative in the high baryon density limit.
    We also disregard the dependence of the field correlators on the magnetic field, 
        which is numerically adequate for \(|eH|\lesssim 0.5\text{ GeV}^2\) as follows from lattice simulations\cite{DElia2015}.

\section{Chiral Vortical Effect in hot QGP}\label{SectCVE}

As a model of a relatively small QGP grain 
    that emerges in HIC
        in which the strangeness density fluctuation is approximately uniform,
            we consider rigidly rotating QCD  in the deconfined phase.
Such a model is consistent with the HIC simulations, 
    where graining (discretization in space and time) is usually larger than 1 fm(/c).
Since the mean free path \(l< T^{-1}\) of a quark\cite{VilenkinCVE,Ambrus2014uqa,Abramchuk2018jhd,Abramchuk2020nrp} in QGP is few times smaller 
    than the typical size of the vortical structure\cite{Zinchenko2022tyg,Csernai2013,Bravina2021arj},
        the rigid rotation may be a satisfactory model for the QGP motion
--- the angular velocity corresponds to the vorticity at any given point inside the fireball.

In ongoing experiments, the vorticity of QGP is at most \(\omega\sim 10^{22}\text{s}^{-1}\sim 10\) MeV \cite{STAR2017ckg},
    which is a relatively small scale 
    (the QCD vacuum correlation length is\cite{Simonov2018cbk} \(\lambda\sim 0.2\text{ fm}\sim 1\text{ GeV}^{-1}\)),
        so we expect the dependence of the field correlators on the angular velocity to be negligible. 
The `kinematic' effect of the gluonic content rotation is non-perturbatively suppressed,
    as was investigated in the Appendix of \refcite{Abramchuk2023}.

The dynamical effect of rotation on the gluonic content of QGP is an intriguing problem on its own.
Rigidly rotating Gluon Plasma was recently reported\cite{Braguta2023yjn} 
    to be unstable at temperatures \(T<1.5T_c\)
        where \(T_c\) is the critical temperature of the Gluon Plasma.

The axial current from the quark propagator \eqref{EqSq} in a rotating frame 
on the rotation axis reads\cite{Abramchuk2023},
    in which we follow the original derivation\cite{VilenkinCVE},
    but in the world-line representation, 
    with the non-perturbative interactions included
    and with all the perturbative corrections neglected
\begin{align}
    \braket{j^5_\mu(x_\bot=0)} \approx &\langle \tr_{c,D} 
        \gamma_5\gamma_\mu (-\slashed D(A, \mu) + m)_{x} 
        \int_0^{+\infty}ds~\zeta(s)~(\overline{\cD^4z})_{xx}^s e^{-m^2s-K}\times\nonumber\\
    &\times P\exp\Big(ig\oint A\cdot dz 
        + \mu\int dz_4 
        +\int_0^sd\tau\sigma^{\rho\sigma}gF_{\rho\sigma}[A](z,z_0)\Big) \times\nonumber\\
    &\times\exp\Big(\frac14\Omega^i\epsilon_{i\rho\sigma 4}\sigma^{\rho\sigma}
        (z_4(s)-z_4(0))\Big)\rangle_{B},
    \label{EqJ5quenched}
\end{align}
where the last exponential is the spin-1/2 rotation operator in the world-line representation. 
With approximations applied in previous sections, we obtain\cite{Abramchuk2023}
\begin{align}
    &\braket{\vec j^5(0)} \approx 
    \vec\Omega~\frac{N_c}{2\pi^2} I_\text{CVE}(T,\mu),\\
    I_\text{CVE}(T,\mu) = -\int_0^{+\infty}p^2&dp
        \Big(f'(\sqrt{p^2+M^2} + V_1/2 -\mu)
        + f'(\mu\to-\mu)\Big),\label{EqICVE}\\
    &\to \frac{\pi^2}{3}T^2 + \mu^2, \text{ at }V_1,M\to 0,\label{EqICVEfree}
\end{align}
again, with the well-known free massless limit\cite{VilenkinCVE}.

\section{Results}\label{SectRes}

\begin{figure}[h]
  \centering
  \begin{minipage}{0.49\textwidth}
    \centering
    \includegraphics[width=\textwidth]{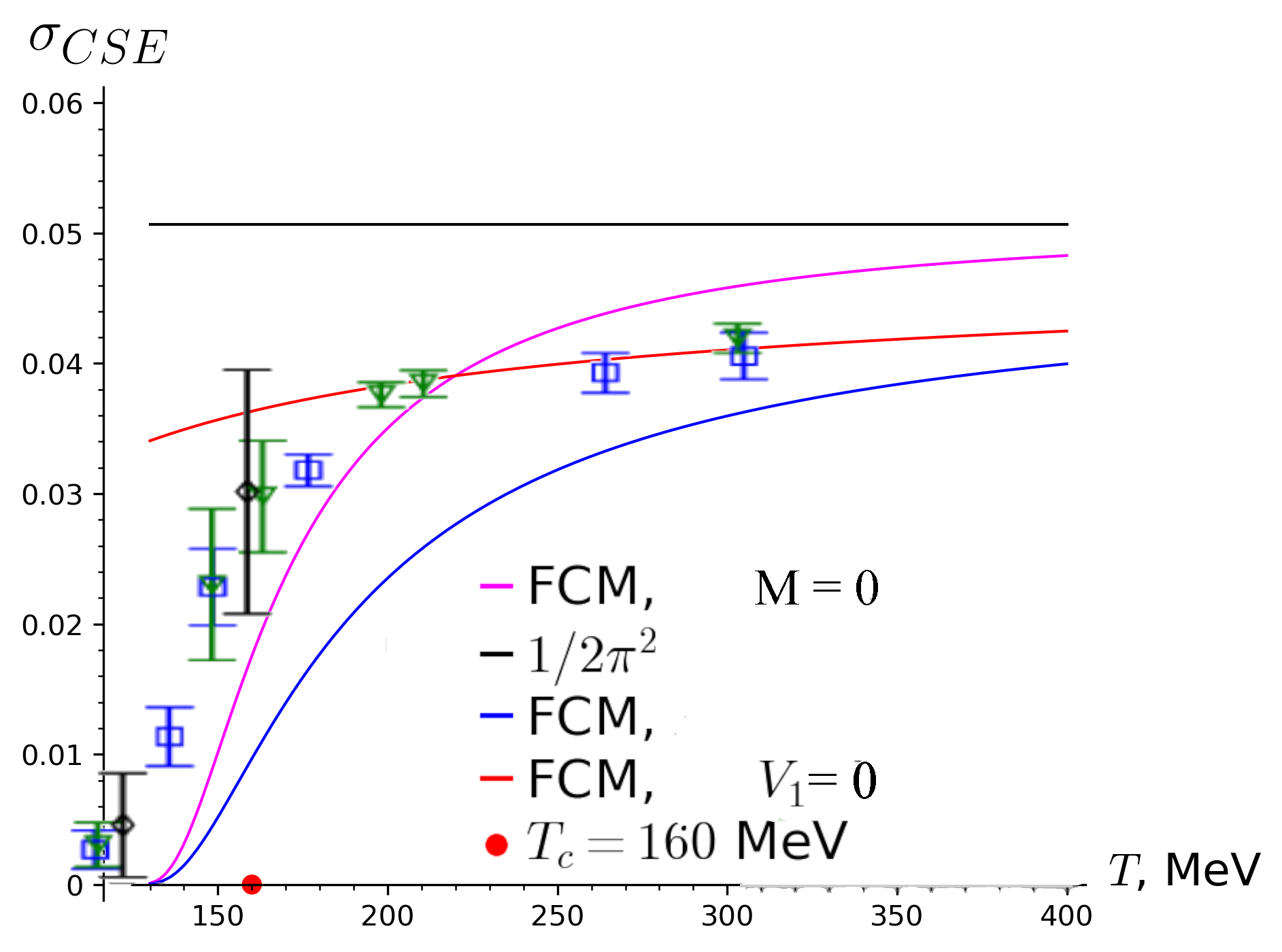}
  \end{minipage}
  \begin{minipage}{0.49\textwidth}
      \centering  
    \includegraphics[width=\textwidth]{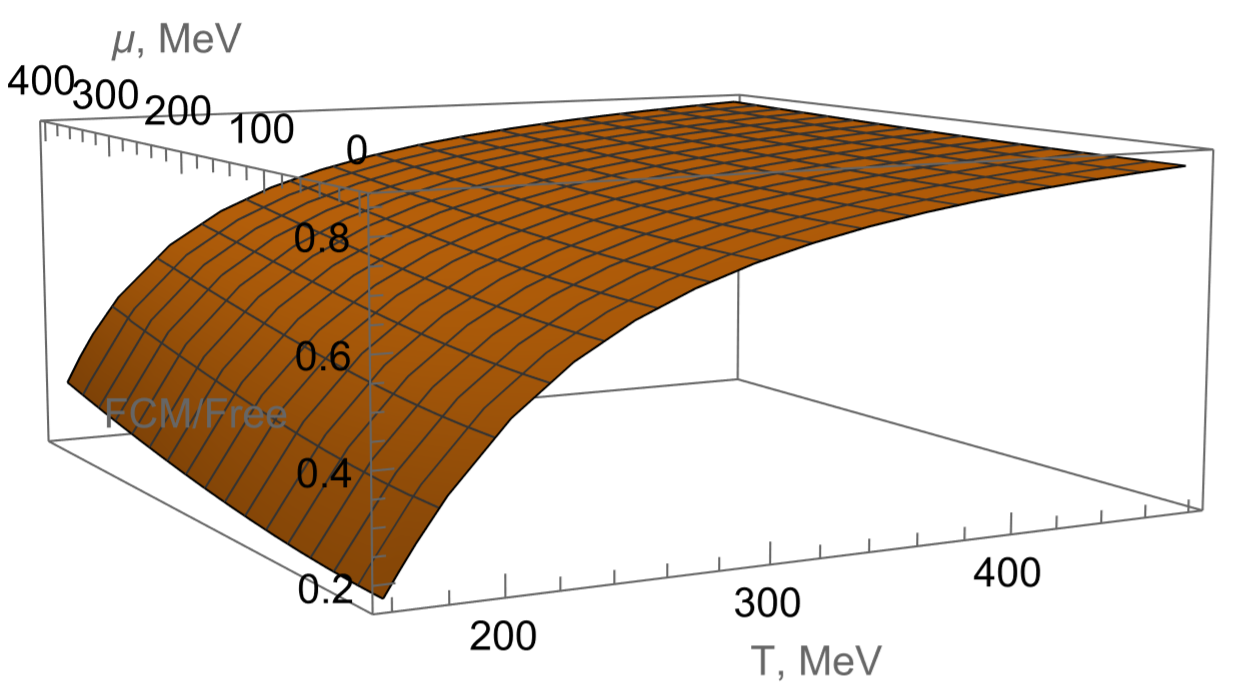}
  \end{minipage}
  \begin{minipage}{0.49\textwidth}
    \centering
      \caption{CSE conductivity \eqref{EqICSE} in the leading non-perturbative order from FCM (the curves) and the free fermions result (the upper const. line), 
      and the lattice data \cite{Brandt2022} (the points).
      }
      \label{FigCSE}
  \end{minipage}
  \hfill\vline\hfill
  \begin{minipage}{0.49\textwidth}
      \centering  
      \caption{Ratio of the CVE axial current in the leading non-perturbative order from FCM \eqref{EqICVE} to the free massless value \eqref{EqICVEfree}.}  
      \label{FigICVE}
  \end{minipage}
\end{figure}

The obtained suppression of CVE is
    in a qualitative agreement with the effect of the phenomenological suppression factor \(k\) in \eqref{EqCvk}
    used in numerical simulations \refcite{Baznat2017jfj} to match RHIC-STAR data. 

Suppressions of CSE and CVE currents 
    (ratio to the free massless limits) 
    are similar,
so we do not insert the figure analogous to Fig.\ref{FigICVE} for CSE.

CSE conductivity comparison (Fig.\ref{FigCSE}) to the lattice data \refcite{Brandt2022,Brandt2023}  hints 
    that the interplay of the non-perturbative color-electric and color-magnetic interactions in FCM deserves further investigation.

The suppression is systematically overestimated,
    since the CMC string forms at distances above the vacuum correlation length 
\(\lambda\).
The same for the approximations 
\(V_1(T,\mu)\approx V_1(T)\), \(\sigma_H(T,\mu)\approx \sigma_H(T)\) ---
    the baryon density might deteriorate the Color-Electric and CMC interactions. 

The present calculations are only for the leading non-perturbative contribution.
Even though the strong coupling in the temperature range of interest might be numerically large 
(\(\alpha_s\sim 0.3\)),
    the non-perturbative background additionally suppresses\cite{Simonov2016xaf} the perturbative corrections 
    with the non-perturbative small parameter \(\sigma\lambda^2\sim1/5\).

A calculation of the leading perturbative correction to CSE current in QED 
    was attempted in  \refcite{Gorbar2013upa}.
The expression for the radiative correction contains infrared divergences. 
Apparently, the non-perturbative interaction (CMC) in QCD would remove the analogous  divergences\cite{Simonov2016xaf}.
However, the calculation with FCM would be even more cumbersome.

\section{Conclusions}\label{SectCon}

We reviewed the calculation of the leading non-perturbative contributions to CVE and CSE for strange quarks in QGP at temperatures above the deconfinement phase transition. 
To deal with the strong interactions of QCD, 
    which are dominant even in the deconfined phase, 
        we used the Field Correlator Method. 

For the experimentally accessible at LHC and RHIC range of temperatures and densities, 
    the Chiral Effects are suppressed by non-perturbative interactions 
    (in comparison to the effects for free fermions),
--- by the Color-Magnetic Confinement,
        which produces a Debye-like screening mass,
        and by the remnants of the Color-Electric interaction,
            which are encoded in the Polyakov line potential. 

The obtained suppression for CVE \eqref{EqICVE} is in a qualitative agreement 
    with the phenomenological suppression (the effect of the factor \(k\) in \eqref{EqCvk}) introduced in the quark-based HIC simulations \refcite{Baznat2017jfj} to match the RHIC-STAR data on the \(\Lambda/\bar\Lambda\)-hyperons spin polarization.
And for CSE \eqref{EqICSE} --- 
    in a qualitative agreement (Fig.\ref{FigCSE}) with the lattice QCD simulations \refcite{Brandt2022,Brandt2023} 
    (a similar result for CSE  with FCM in strong magnetic field limit was also obtained in \refcite{Khaidukov2023pfu}).
Incorporation of our results into the quark-based numerical simulations of HIC would provide an ultimate check.

\section*{Acknowledgements}

The results reviewed in the present paper have been obtained in collaboration with M.A.Zubkov and M.Selch.
The author is grateful to A.S.Sorin for useful discussions.

%
%
%

\bibliographystyle{ws-ijmpe}
\bibliography{CEinHIC}

\end{document}